\documentclass[10pt]{article}
\usepackage{setspace}
\begin{document}
\title{Crossing the Line: Towards increasingly fruitful complex systems research for the physics community}
\author{Reginald D. Smith \\ PO Box 10051, Rochester, NY 14610 \\ rsmith@bouchet-franklin.org}
\date{September 22, 2011}

\maketitle
\begin{abstract}
This article addresses broad trends in interdisciplinary work in complex systems from the physics perspective where interactions with colleagues in fields such as computer science, ecology, or economics can often be derailed by unintentional clashes of methodologies and perspectives on the core science. Key causes of such breakdowns in interdisciplinary work are detailed and solutions offered.
\end{abstract}
\newpage 

\section{Introduction} 

In many ways, the state science represents the vanguard of human knowledge, but in one meaningful way it is reclaiming old ways. In the beginnings of modern scientific research, the current rigid divisions between disciplines were nearly nonexistent and the blanket term ``natural philosophy'' was applied to almost all efforts. The more well-known and professional researchers gathered under the auspices of scientific societies such as the French Academy of Sciences and Britain's Royal Society. Over time, starting in the 19th century, as scientific knowledge increased, science became more professionalized, grants became the primary vehicle for funding, the moniker natural philosophy began to disappear and science began to establish many specific disciplines such as physics, chemistry, biology, and geology. Over the 20th century, these distinctions became even more refined and numerous sub-disciplines and related journals sprang up. 

As we begin the 21st century, the barriers that have separated many of these disparate fields for over a century are falling and previously unrelated areas such as physics, biology, polymer chemistry, and even linguistics are beginning to see fruitful cross-disciplinary collaboration and new results. It is a welcome development and has breathed life into all fields that participate. Physicists, because of strong reductionist training and well-developed techniques for dealing with complex systems in a quantitative fashion, have found themselves in a prime position to take advantage of these trends. A prime example is complex network theory which has combined earlier results from mathematicians and sociologists with statistical mechanics and percolation theory to investigate the architecture of everything from computer networks to cellular metabolism. 

There is also a deeper history to the involvement of physicists in other disciplines. In the 1930s many logical positivist philosophers, chief among them Otto Neurath, embarked on an effort to unify the sciences \cite{positivist}. Though their efforts largely focused on philosophical aspects it was commonly agreed that physics was the root of all other disciplines which were seen as supervening on underlying physical principles. In other words their properties were based on lower-level physical phenomena. This radical reductionism has proved difficult in practice and no one expects psychology to be reduced to mean-field analysis anytime soon. In fact, most research by physicists in other disciplines seeks not to reduce the systems to fundamental physical variables but to apply well-developed techniques from statistical physics to these systems. The unity of science efforts though not well-known, influenced no less than Thomas Kuhn and physics' fundamental position is widely recognized in science education. 

On the other hand, this perspective has irritated some of our colleagues in other disciplines who often, for both right and wrong reasons, call physicists arrogant and sometimes resent what they believe is an 'intrusion' on their discipline. The eminent ecologist Lawrence Slobodkin was completely serious when he accused scientists from other disciplines, presumably mathematics and physics, of "carpetbagging" their way into ecology \cite{slobodkin}. Interdisciplinary research will undoubtedly not only continue but increase. Interdisciplinary research, like all research however, carries not only huge promises but hidden perils. In this paper, I will not discuss strategies for forming collaborations or analyze past efforts but rather focus specifically on what I see as a great need for improvement in much of the interdisciplinary work both physicists and mathematicians are doing. These concerns have arisen both from my my experience publishing papers using techniques from physics for linguistics \cite{reggie1,reggie2}, econophysics \cite{reggie3}, plant-mycorrhiza mutualisms \cite{reggie4}, and Internet traffic \cite{reggie5} but also from discussions with experts in these disciplines on their perception of the work of physicists in their fields and what has been done right and wrong. My intention is not to single out any field of collaboration as being more guilty but to give general guidance on how physicists can make interdisciplinary research more accessible, useful, and respected by our colleagues in other disciplines. 

\section{Promises and perils of interdisciplinary research} 

The promises of interdisciplinary research have become so well-known, they are commonly held as platitudes. One vivid recent example was the 2009 Nobel Prize in Chemistry where Venkatraman Ramakrishnan received his PhD in physics and later worked on the problem of ribosome function. Interdisciplinary research has provided new perspectives and led to the development of valuable tools in physics such as the Kuramoto coupled oscillator model, scale-free network topologies, and quantitative models of biological systems. Usually the most prominent contributions of physicists are the introduction of quantitative mathematical models of well-understood systems that reproduce known dynamics and predict new dynamics under certain conditions. In mathematical ecology, the many permutations of the predator-prey competition equations is one of the most well-known examples. 

However, one would have to be honest that not all such forays into new territory are as fruitful and useful as many of us originally hope. There can be many reasons for this but I will try to outline some of the most prominent. 

\subsection{Multidisciplinary versus true interdisciplinary research} 

The distinction of a multidisciplinary versus an interdisciplinary effort is well-known amongst organizational behavioral analysts who study group collaboration. In short, in a multidisciplinary setting everyone brings their own perspective to the research topic at hand but there is not an united and cohesive effort to fit all the ideas and results into a coherent whole. In true interdisciplinary research, there is not only a contribution by each group but a joint effort to synthesize these efforts into a framework that is both coherent and useful in advancing the knowledge and state of the art for all parties.  

The truth is that a lot of what we think has been interdisciplinary work in the physics community has unfortunately instead been firmly multidisciplinary. Often where we can have groups of physicists and other colleagues attacking a problem jointly, sometimes we just have physicists attacking the problem in the other discipline without feedback or reverence for the professionals in the other discipline. This can form a useful purpose of introducing innovations and bypassing over-conservative and slower moving orthodox trends. However, every paper in a new field is not necessarily a breakthrough. Therefore, we oftentimes see physicists reinventing the wheel and claiming new insights on already understood phenomena without integrating it with the current state of the field. 

What is needed is not for physicists to "bow down" or only follow current perspectives in a field but to be cognizant on how their new insights fit into the greater whole. Applying a successful model from statistical mechanics to a many-body complex phenomenon in a biological or ecological system can be helpful but only as far as the models reflect reality as well as integrate and predict known insights in the discipline in question.

\subsection{Parallel research tracks on the same problem} 

A key outgrowth of the first issue is that different fields working on the same problem, i.e. tumor angiogenesis or Internet traffic self-similar behavior, develop two different tracks of research from increasingly different angles. Neither angle is "right" in the sense of one being useful and the other useless, however, over time these approaches can, like languages or population genetics, widely diverge to the point of being mutually unintelligible. For example,  since a seminal paper by network engineers Leland, Willinger, \& Wilson and mathematician Taqqu \cite{selfsimilar} in 1994, it has been widely recognized that time series of Internet traffic in numbers of packets displays both self-similarity (extremely large variances and bursty behavior at all timescales versus typical Poisson distributed behavior) and long-range dependence (a non-finite summation over the values of an infinite time lag autocorrelation output). The bulk of the research has been done by engineers attempting to understand the nature of network traffic to construct better and more reliable network technologies. However, with a similar experience dealing with self-similarity in physical systems, physicists soon also published papers commenting and offering a different perspective. Where network engineers determined self-similarity was largely caused by long-tail distributions in the sizes of files on the Internet, physicists adapted tools from network topology and critical phenomena to reproduce self-similar behavior in models. 

From here arose two issues that are the heart of the problem of parallel research tracks. First, diverging lines of inquiry. Though the first papers by physicists seemed aware of the nuances of the work by network engineers and quoted their work in IEEE and ACM respectfully, over time the research in physics diverged into investigating theoretical network models on different topologies and with different traffic strategies. Soon they no longer quoted any of the relevant network engineering literature on the problem, besides the initial 1994 paper, and unfortunately became disconnected from actual measurement of Internet traffic and its behavior. This point was realized by network engineers evaluating the work, who began to largely disdain such lines of inquiry and unfortunately did not engage physicists in improved research \cite{willinger1, willinger2}. 

In particular, even amongst the sympathetic and interested members of the network engineering community, they raised a pertinent complaint which is the second main cause of parallel research tracks. Namely, lack of validation of models with the system of study. Cross-discipline lack of communication can cause divergent paths of research but one hopes with regular testing and validating of theories against data from the system of study, the paths will never diverge too far and may even converge together down the road\cite{caida}. However, the very nature of the research methods being used by both groups makes this difficult. Physicists try to look at a  picture of overall network topology and packet traffic but use simplistic topologies such as 2D lattices and computer traffic models which are rarely if ever validated against the real structure and data from Internet traffic traces. Network engineers made crucial steps in validating the file-size distribution causes of self-similarity by analyzing actual file size distributions on the Internet and comparing them to theory. They also know the subtle nuances of analyzing network traffic. However, it is extremely difficult for both technical and proprietary reasons to obtain traffic traces over multiple Internet nodes simultaneously and thus makes any analysis to confirm or refute topological effects on data traffic exceedingly hard.  

However, there needs to be an effort to reconcile models with system data, no matter how difficult. Willinger and collaborators stated it well in \cite{willinger1,willinger2} where they emphasized the need to ``close the loop'' validating models against data and not use na\"{i}ve statistical mechanics models. I would not go so far as to completely dismiss the research physicists have done and what they can contribute in the future but ``closing the loop'' is a good lesson for us to adopt in this or any other interdisciplinary research problem.

\subsection{Theoretically beautiful but not realistically viable} 

A problem related to parallel research tracks ironically derives from one of the strengths of physics. Physics has many well-researched and theoretically elegant models to investigate many different types of behavior. Even phenomena thought to be mind numbingly complex such as phase transitions or ferromagnetic domain cascades in hysteresis are accurately described by these models. Therefore, when faced with a problem where we have a choice to maim the beautiful model to fit an ugly world or maim our representation of the world to fit a beautiful model, we choose the latter. This particularly becomes a problem when we deal with systems with non-ideal components such as ecological food webs, disease carriers in epidemiology, or the dynamics of social insects such as many of the Hymenoptera.  

Physics has come a long way since experimental luminaries such as Ernst Mach brutally chastised Ludwig Boltzmann for basing his theories off atoms asking, ``have you seen one?'' In fact, the trailblazers of quantum mechanics were sometimes awful experimenters (Wolfgang Pauli) or pure theoreticians. Therefore, physics, for good reasons, has come to prize concise and elegant theoretical insight as a useful tool. But just like a toolbox has both hammers and wrenches, pure theoretical derivations may not always be the appropriate way to attack a problem. 

When we try to apply a beautiful theory to a system composed of non-ideal components several problems can arise. First, since the theory is an artifact of another scientific problem it was imported from it may more resemble a black box solution than a bottom-up modeling of actual behavior. Second, theoretical models in non-ideal situations can make the actual value or utility of their variables fuzzy. Therefore, complex and multi-stage growth and metabolism are reduced to Malthusian growth constants, variables are included in the model which are really more exercises in curve fitting than actual relation to the system, and the values of variables can be arbitrary and made to fit data without trying to emphasize or explain their actual value in the context of the system dynamics. Some of the variables introduced have no relevance to the system outside of the theoretical realm or could never possibly be measured with any sort of precision. 

Third, in dealing with complex systems we often deal with complex feedback loops among multiple parts of the system. However, often this feedback is simplified or eliminated altogether to develop an open loop equation in the name of analytical tractability.  Reducing a system of equations can work in approximation if the correlation between distinct parts is included in modeling the dynamics but this is not always done. Finally, complex systems are by definition open systems and therefore modeling them while neglecting environmental variables or effects could understate feedback amongst environmental variables external to the system and creates a false reductionism which breaks the house down to its individual bricks but completely ignores the foundation. 

This can also link back to the previous problem of lack of validation with real systems. Key examples are the increasingly complex permutations of predator-prey relations and their stable solutions related to the Lotka-Volterra model. The Lotka-Volterra predator-prey model is extremely important and useful as a model of real species dynamics. Granted, however, even in its simplest two-species model incorporating logistic growth-type population constraints, the various demonstrations of oscillations and chaotic population dynamics have been difficult to verify with large amounts of empirical data. Part of this is an artifact of the difficulty of defining the range for a species and accurately making a census of its population while neglecting immigration and other impacts. However, this has not stopped researchers from developing ever more elegant, complex, N-species models whose possibilities of validation against real dynamics can be increasingly tenuous.  

In an excellent primer on ecological modeling \cite{weiner}, Jacob Weiner demonstrates that sometimes the relationships between real ecology and models can be lacking. He excellently emphasizes that one should remember "the model is not the object of study" and more succinctly uses an analogy for the problems modelers often run into \cite{weinercomm}:

\begin{quotation}
the emphasis on mathematical elegance and rigor among modelers is like a football coach saying -``Wouldn't it be beautiful if the players' movements on the playing field were synchronized?'' Yes, it might look very nice, but the team would not win many games.  Visually it might be great, but it would be terrible as football.\end{quotation} 

Ironically, one of the most maligned groups of modelers in science, climatologists, have probably proved ahead of many other disciplines in trying to validate their models against measured data from a variety of sources documenting historical temperature trends. In large complex systems, sometimes models may be the only realistic way to proceed and given this one may be inclined to sacrifice elegance for realism in order to present a theory that both reproduces measured data and also proves useful to those in other fields to model and predict data according to their own methodologies.  

Given this necessity, it is a bit ironic that climatologists and economists are often at each others throat since they both confront the same problem from a different perspective. Both human economies and the global climate are large-scale non-equilibrium complex systems with vast numbers of variables. Whatever difficulties we may have forecasting temperature in even the next year could be said about forecasting GDP growth or equity market performance. Perhaps in a better world these two groups would collaborate to enrich our overall knowledge of modeling and forecasting complex systems but maybe it is true in part that familiarity breeds contempt.

\subsection{Incorrect conclusions from general phenomena} 

The final issue with interdisciplinary research I want to discuss is a frequent error of deriving incorrect conclusions from general features of complex systems. Oftentimes a model is deemed `successful' if it superficially generates aspects of the system in question. Because these aspects are simulated, it is felt that validation is unnecessary or even unwanted since empirical observation may be unable to reproduce the precision of the simulated output.  

One of the biggest areas where this seems to occur is where systems feature some sort of power law or fractal description. Many systems feature either time series that are self-similar, display $1/f$ noise in spectral decomposition, or have topological features which can be described as fractal. Fractal dimension in particular is oftentimes only used in a descriptive fashion without links to the generative dynamics of the system or an explanation how the fractal dimension can explain or predict system dynamics. Self-similarity appears in too many different systems and under too many different conditions to be completely ignored as a coincidence. However, though the general behavior is similar, there are large classes of dynamics that can reproduce such behavior. This is the hardest criticism for me to write since I see so much promise in this line of research. When a complex system shows a change and behavior past a critical value for a variable and even displays critical slowing down near the critical point, it seems to beg for an explanation on high from the theory of phase transitions. It reproduces the main aspects of such systems and is likely not a coincidence. 

However, though this may be true we have to peel under the first layer of observation of fractals/self-similarity to truly understand the system. We discovered this in complex networks when it was realized that not just a power law degree distribution but the distribution of connections between nodes by degree (assortativity) is vital in describing a network. Several papers that should be required reading amongst all researchers finding power laws in their data are Clauset, Shalizi, \& Newman \cite{powerlaw1} who discuss the accurate and rigorous measurement of power law phenomena and Mitzenmacher \cite{powerlaw2,powerlaw3} who discusses generative models for power law behavior. Additional constructive criticism on scale-free networks is given by Li et. al. \cite{powerlaw4}. Also Ferrer i Cancho and Sol\'{e} in their paper \cite{zipf} on stochastic generation of Zipf's Law in texts shows that the generation of Zipf's Law through the ``monkeys on typewriters'' analogy of random words is seriously flawed--the inverse Zipf plot where the probability of a word having a frequency of appearance in the text $P(f)$ is plotted vs. frequency shows non-power law behavior in contrast to the power law behavior of human languages. 

From the linguistics perspective, a sympathetic yet critical voice is Gabriel Altmann, one of the foremost experts on Zipf's Law. In two papers \cite{altmann1,altmann2} he specifically addresses the entry of physicists in linguistics in a nuanced way, inviting their participation in research but warning against naïve theories which try to explain certain patterns in linguistics by trying to force analogies to physical concepts (like statistical or quantum mechanics) which generate superficially similar behavior. He also highlights the genuine contribution to linguistics by by Ioan-Iovitz Popescu as an example of how an outsider can help advance the field by creating his own ideas and communicating them directly with linguists. Popescu's research \cite{popescu} of the h-point on the rank-frequency curve of words, where the power law behavior collapses for words of relatively high-rank and begins a rapid decline, has helped enrich our knowledge of the limitations of the pure Zipf's Law (the zeta distribution).

\section{Conclusion} 

In this paper, I hope I have not made an impression that I want less interdisciplinary work from physicists, quite the opposite. I also don't purport to judge all previous work since my insights in this paper stem mainly not from criticism of others but from honest feedback and appraisal of some of my own results and past papers.  

I believe in the theory of complex systems there is some sort of discovery on the horizon. Too many things in too many disciplines are converging for it to be otherwise. However, to advance to the next level we must evaluate our present state and research. There is much to be done, however, it will take truly interdisciplinary work that does not just force the results from one field onto another but consolidates observation and sound theory into a firm an coherent whole that guides future results and research. I think the physics community as a whole is more than prepared for its task and therefore it is my hope that this paper is not a lament of a permanent decline in standards but rather a small helping hand for a future revolution of knowledge we have yet to uncover.

\end{document}